\documentclass{article}
\usepackage{ijcai26}

\usepackage{times}
\usepackage{soul}
\usepackage{url}
\usepackage[hidelinks]{hyperref}
\usepackage[utf8]{inputenc}
\usepackage[small]{caption}
\usepackage{graphicx}
\usepackage{amsmath}
\usepackage{amsthm}
\usepackage{booktabs}
\usepackage{algorithm}
\usepackage{algorithmic}
\usepackage[switch]{lineno}

\usepackage{amsfonts}
\usepackage{multirow}
\usepackage{adjustbox}
\usepackage{amssymb}
\usepackage[breakable]{tcolorbox}
\usepackage[table]{xcolor}
\usepackage{pifont}

\title{Structure-Guided Memory Consolidation for Mitigating Compounding Errors in Literature Review Generation}

\author{
Zhi Zhang$^{1}$\and
Yan Liu$^{1,*}$\and
Zhejing Hu$^{1}$\and
Gong Chen$^{2}$\and
Shenghua Zhong$^{3}$\and
Sean Fontaine$^{2}$\And
Jiannong Cao$^{1}$\\
\affiliations
$^{1}$The Hong Kong Polytechnic University, Hong Kong, China\\
$^{2}$Biaohan Limited, Shenzhen, China\\
$^{3}$Shenzhen University, Shenzhen, China\\
\emails
\{zhi271.zhang, zhejing.hu\}@connect.polyu.hk,
\{yan.liu, jiannong.cao\}@polyu.edu.hk,\\
heinz.g.chen@hotmail.com,
csshzhong@szu.edu.cn,
sean@clozzz.com
}

\begin{document}

\maketitle

\renewcommand{\thefootnote}{}
\footnotetext{*Corresponding author}

\begin{abstract}
Compounding errors pose a significant challenge in automatic literature review generation, as inaccuracies can cascade across multi-stage retrieval and generation workflows. Existing self-correction strategies often lack mechanisms to effectively track and consolidate verified information throughout the process, making it difficult to prevent error accumulation and propagation. In this paper, we propose Structure-Guided Memory Consolidation (SGMC), a novel framework that incrementally consolidates and verifies information using structured representations at each stage of the literature review pipeline. SGMC consists of three key modules: Tree-Guided Memory for hierarchical literature retrieval and outline generation, Hub-Guided Memory for evidence extraction and iterative content refinement, and Self-Loop Memory for proactive error correction via historical feedback. Extensive experiments on public benchmarks and a newly constructed large-scale dataset demonstrate that SGMC achieves state-of-the-art performance in citation accuracy and content quality, significantly mitigating compounding errors in long-form literature review generation.
\end{abstract}

\section{Introduction}
Literature reviews play an important role in scientific research by synthesizing existing knowledge to identify research gaps, establish theoretical frameworks, and guide future studies \cite{ermel2021literature,liu2023generating}. Over recent decades, the rapid growth in the number of scientific publications has made it increasingly challenging for researchers to efficiently retrieve, organize, and synthesize relevant information in their fields \cite{wang2024autosurvey}. In response, automatic literature review systems that leverage advances in artificial intelligence have become increasingly valuable, as they help reduce manual workload, provide timely coverage of new research, and facilitate the identification of key trends and research gaps within the extensive literature \cite{zhang2025mixture,wang2024autosurvey,liu2023generating}.

Early approaches primarily leverage multi-document summarization techniques to generate summaries from a set of reference papers \cite{hoang2010towards,hu2014automatic,erera2019summarization}, enabling the automatic creation of related work sections given a set of references as input. In recent years, large language models (LLMs) have demonstrated impressive planning and reasoning abilities \cite{wang2024survey,zhao2024expel,zhong2024can,nandy2025language,zhang2025mixture}. This has led to a growing research area that employs LLMs as controllers to construct autonomous agents capable of effectively performing complex tasks \cite{wang2024survey}. Building on these advances, recent state-of-the-art research has achieved remarkable success by designing LLM-based autonomous agents to automate the entire pipeline of literature review and generate long-form manuscripts \cite{wang2024autosurvey,liang2025surveyx}.

\begin{figure*}[h]
    \begin{center}
        \includegraphics[width=\linewidth]{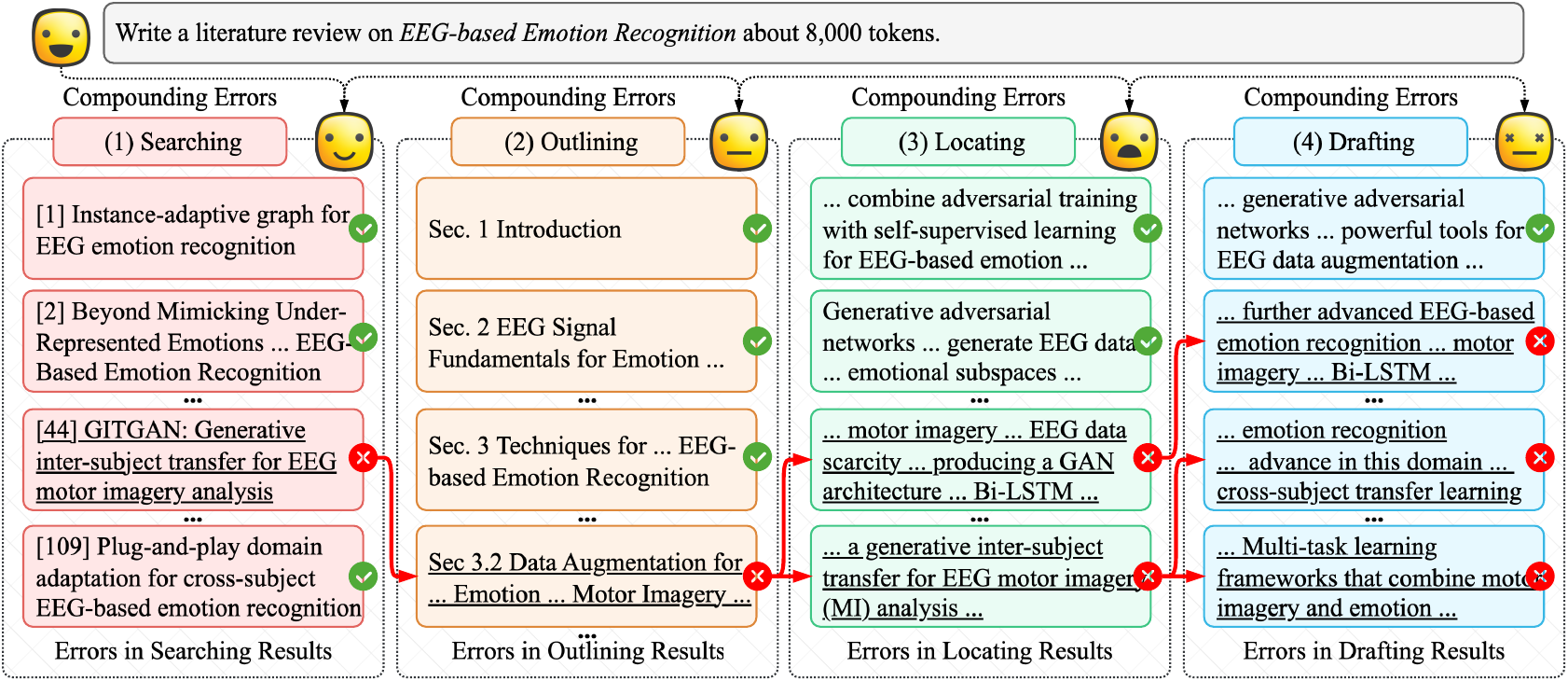}
    \end{center}
    \caption{Illustration of the error accumulation phenomenon.}
    \label{fig:paradigm}
\end{figure*}

Although these advances enable the automatic generation of human-like literature reviews with reduced manual effort, previous studies on multi-step models show that the inherent long-horizon workflow of such systems inevitably introduces the risk of compounding errors. In the workflow, small mistakes made at earlier steps can accumulate and propagate through subsequent steps, as errors that occur at a certain step affect the next state \cite{wang2025expressive,han2023expert,venkatraman2015improving,hejna2023improving}. This paper investigates the phenomenon of error accumulation in automatic literature review generation. As a demonstration, we employ the large language model GPT-4.1 \cite{openai2025gpt41} to conduct a literature review generation workflow on the topic of \textit{EEG (electroencephalogram)-based emotion recognition}, with a target length of 4,000 words. As shown in Fig.~\ref{fig:paradigm}, when the retrieval stage fails to identify relevant literature, the resulting errors can mislead the outlining process. An irrelevant outline may further cause the drafted manuscript to deviate from the intended topic.

However, directly applying these methods faces challenges in the context of literature review generation. Applying self-correction to the final output is difficult because errors have already accumulated across multiple stages, making the result hard to repair. Applying self-correction at each intermediate stage fixes errors within that stage but neglects context across stages. Thus, we propose to construct a memory mechanism that consolidates verified information along the long-horizon workflow using structured representations. Results of stages in the workflow are verified and consolidated into the memory, and subsequent steps expand from this foundation. Errors along the long horizon are filtered during consolidation and replaced during expanding before they can propagate, narrowing the search space around verified facts.

Based on these observations, we propose the Structure-Guided Memory Consolidation (SGMC) framework for automatic literature review generation. Specifically, Tree-Guided Memory ensures each outline level is grounded in actual literature, reducing topic drift. Hub-Guided Memory tightly aligns draft content with supporting evidence through iterative refinement. Self-Loop Memory introduces a dynamic checklist mechanism, enabling agents to learn from historical errors and proactively prevent error propagation.

\begin{figure*}[t]
    \begin{center}
        \includegraphics[width=\linewidth]{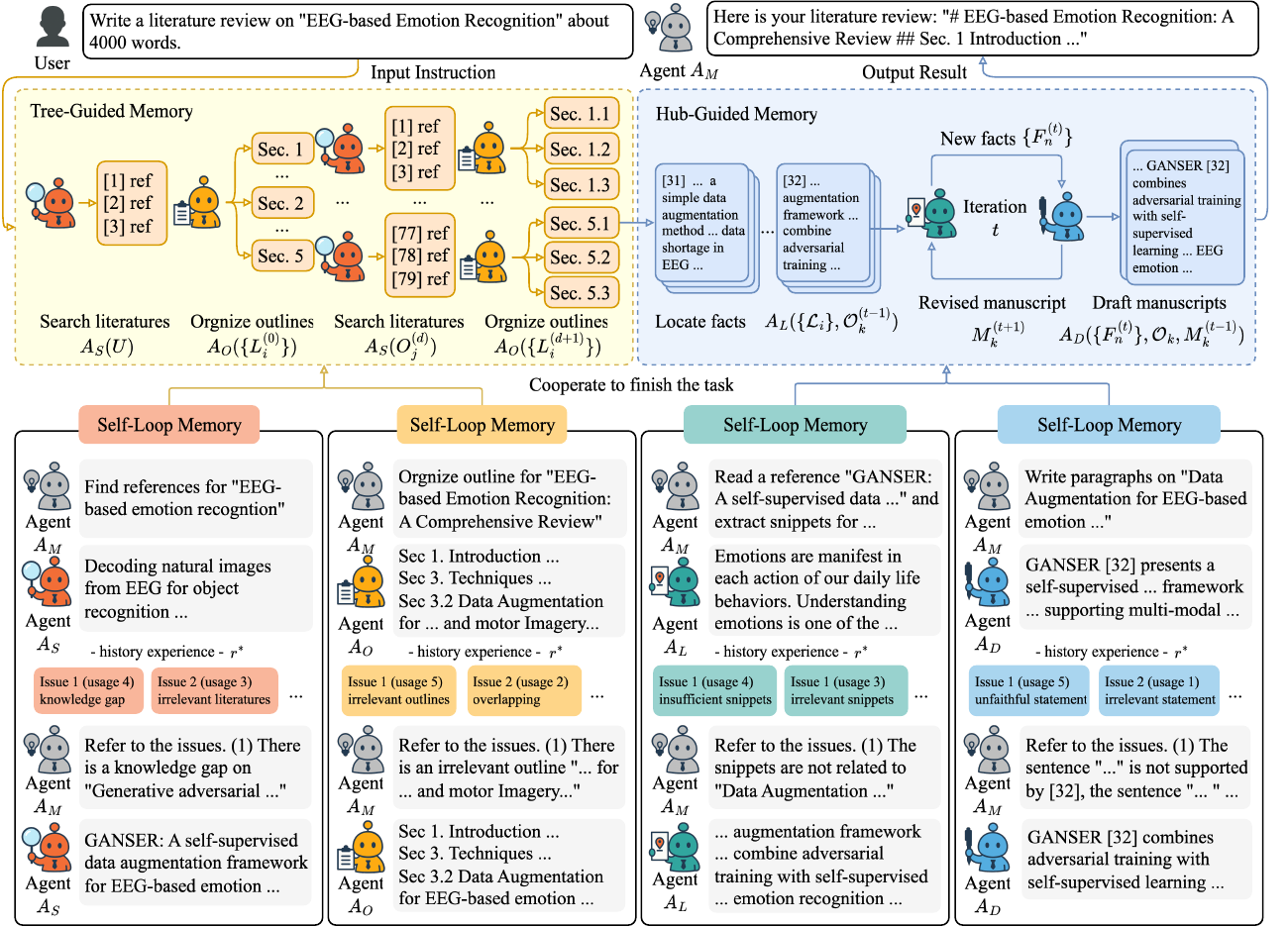}
    \end{center}
    \caption{The overall framework of Structure-Guided Memory Consolidation (SGMC) for literature review generation.}
    \label{fig:framework}
\end{figure*}

\section{Related Work}

Automating the summarization of research papers has long attracted interest. Early work used multi-document summarization techniques. In 2010, Hoang \textit{et al.} introduced automatic related work summarization, extracting sentences from multiple articles to form an extractive summary \cite{hoang2010towards}. Lu \textit{et al.} constructed Multi-XScience, a dataset for generating related work sections based on a paper's abstract and cited articles, which became a benchmark for later studies \cite{lu2020multi}. Chen \textit{et al.} proposed a relation-aware multi-document encoder and a relation graph to capture cross-document dependencies for related work generation \cite{chen2021capturing}. Xiao \textit{et al.} developed PRIMERA, a pre-trained encoder-decoder model with objectives for linking and aggregating information across documents \cite{xiao2022primera}. Liu \textit{et al.} proposed aligning sentences by category and using sparse transformers to organize information from references \cite{liu2023generating}.

Recent advances in LLMs have improved the scope, enabling end-to-end systems to generate literature review papers. Wang \textit{et al.} introduced AutoSurvey, a framework with retrieval, outline generation, parallel subsection drafting, integration, and evaluation \cite{wang2024autosurvey}. Shao \textit{et al.} developed STORM, which models the pre-writing stage by discovering diverse perspectives, simulating expert conversations, and curating information to create an outline \cite{shao2024assisting}. Liang \textit{et al.} designed SurveyX, an end-to-end solution for automated survey generation, covering online literature search, organization, and survey writing \cite{liang2025surveyx}. However, little attention has been paid to mitigating the risk of errors in generated outputs.

A response is post-hoc self-correction and refinement. Self-Refine \cite{madaan2023self} uses a single LLM as the generator, refiner, and feedback provider, iteratively improving outputs through self-generated feedback. Reflexion \cite{shinn2023reflexion} reinforces language agents through linguistic feedback, where agents reflect on task feedback signals and maintain reflective text in an episodic memory buffer to induce better decision-making in subsequent trials. ReConcile \cite{chen2024reconcile} enhances collaborative reasoning via multiple rounds of discussion among diverse LLM agents, employing a confidence-weighted voting mechanism to reach consensus. Chain-of-Note \cite{yu2024chain} generates sequential reading notes for each retrieved document, enabling evaluation of their relevance before formulating the answer. RARR \cite{gao2023rarr} finds attribution for the output of any text generation model and post-edits the output to fix unsupported content while preserving the original output. However, in long-horizon literature review generation, accumulated errors may propagate too far for effective post-hoc repair, and directly applying these methods fails to provide reliable attribution and revision.

\section{Methodology}

As illustrated in Fig.~\ref{fig:framework}, given a user instruction $U$, the proposed Structure-Guided Memory Consolidation (SGMC) framework orchestrates three memory structures to produce the final literature review output $Z$.

\subsection{Tree-Guided Memory}

The Tree-Guided Memory is designed for the stage of outlining the literature review and retrieving relevant references for each section of the outline. Without prior knowledge of the relevant references, the initial outline tends to be suboptimal. Conversely, without a determined direction, the literature retrieval process lacks focus. To mitigate compounding errors, we propose a hierarchical exploration strategy.

In detail, given a user instruction $U$, the agent $A_M$ based on a large language model is responsible for coordinating sub-agents using a Tree-Guided Memory. First, $A_M$ constructs a tree with the user instruction as the root node, representing the overall direction for exploring the field. This tree is expanded by identifying sub-directions of interest. For the root node, $A_M$ first invokes the agent $A_S$, which is responsible for constructing search queries and then calling search tools to retrieve references based on the user instruction, resulting in a set of relevant papers:
\begin{equation}
    \{L^{(0)}_1, L^{(0)}_2, \dots, L^{(0)}_I\} = A_S(U)
\end{equation}
where $L_i^{(0)}$ denotes the $i$-th piece of reference retrieved for the root node at tree depth $d=0$, and $I$ denotes the number of references retrieved at the root level.

Next, $A_M$ assigns the agent $A_O$, which is responsible for outlining, to determine the sub-outline based on the titles and abstracts of the retrieved literature:
\begin{equation}
    \{O_1^{(0)}, O_2^{(0)}, \ldots, O_J^{(0)}\} = A_O(\{L^{(0)}_i\}_{i=1}^{I})
\end{equation}
where $O_j^{(0)}$ denotes the $j$-th sub-outline at the root level, and $J$ is the total number of sub-outlines at tree depth $d=0$.

For each sub-outline node $O_j^{(d)}$ at depth $d$, $A_M$ decides whether further decomposition is needed according to the following criterion:
\begin{equation}
    \phi(O_j^{(d)}) = 
    \begin{cases}
        1, & \text{if } \text{complexity}(O_j^{(d)}) > \theta \\
        0, & \text{otherwise}
    \end{cases}
\end{equation}
where $\text{complexity}(O_j^{(d)})$ denotes the complexity of direction $O_j^{(d)}$, and $\theta$ is the complexity threshold. In this work, we use a simple yet effective strategy in which each direction is treated as a section, and complexity is measured by the number of words.

For outline nodes requiring further decomposition, where $\phi(O_j^{(d)}) = 1$, the process is applied recursively as follows:
\begin{equation}
    \begin{aligned}
        \{L^{(d+1)}_1, L^{(d+1)}_2, \dots, L^{(d+1)}_{I'}\} &= A_S(O_j^{(d)}) \\
        \{O_1^{(d+1)}, O_2^{(d+1)}, \ldots, O_{J'}^{(d+1)}\} &= A_O(\{L_i^{(d+1)}\}_{i=1}^{I'})
    \end{aligned}
\end{equation}
where $I'$ denotes the number of literature pieces retrieved at depth $d+1$, and $J'$ denotes the number of sub-directions at depth $d+1$.

This process constructs child nodes in the tree $\mathcal{T}$, where nodes $\{O_1^{(d+1)}, O_2^{(d+1)}, \ldots, O_{J'}^{(d+1)}\}$ are the children of node $O_j^{(d)}$. The exploration process terminates when the following condition is met:
\begin{equation}
    \forall v \in \mathcal{V}: \phi(v) = 0 \quad \text{or} \quad d(v) = d_{\max}
\end{equation}
where $\mathcal{V}$ denotes all leaf nodes in the outline tree, $d(v)$ denotes the depth of node $v$ in the tree $\mathcal{T}$, and $d_{\max}$ is the maximum allowed depth. The resulting tree $\mathcal{T}$ provides a hierarchical outline, with associated literature at each level, for downstream generation.

\subsection{Hub-Guided Memory}

The Hub-Guided Memory is designed to extract relevant snippets from the literature and draft the review manuscript. Without clear target content, it is difficult to locate the necessary evidence in the literature. Conversely, without sufficient supporting evidence, the manuscript may lack grounding. To mitigate compounding errors, we propose an iterative exploitation strategy.

In detail, given the outline tree $\mathcal{T}$, $A_M$ initiates the Hub-Guided Memory. For each leaf node in the tree, $A_M$ identifies its parent outline node, denoted as $\mathcal{O}_k$, and the associated reference set $\{\mathcal{L}_i\}_{i=1}^{I_k}$, where $I_k$ is the number of references associated with outline node $\mathcal{O}_k$, connected to the hub node. Then, $A_M$ coordinates sub-agents to aggregate information from references to the hub node through iterative cycles. At the initial iteration $t=0$, the agent $A_L$, which is responsible for locating, extracts relevant information based on the outline node $\mathcal{O}_k$ and the literature set $\{\mathcal{L}_i\}_{i=1}^{I_k}$:
\begin{equation}
    \{F_1^{(0)}, F_2^{(0)}, \ldots, F_{N}^{(0)}\} = A_L(\{\mathcal{L}_i\}_{i=1}^{I_k}, \mathcal{O}_k)
\end{equation}
where $F_{n}^{(0)}$ denotes the $n$-th note snippet extracted from the reference set $\{\mathcal{L}_i\}_{i=1}^{I_k}$ relevant to the outline node $\mathcal{O}_k$, and $N$ is the total number of snippets.

The agent $A_D$, which is responsible for drafting, composes the initial draft as follows:
\begin{equation}
    M^{(0)}_k = A_D(\{F^{(0)}_n\}_{n=1}^{N}, \mathcal{O}_k)
\end{equation}
where $M^{(0)}_k$ represents the initial manuscript for the outline node $\mathcal{O}_k$, incorporating the content of all its child outline nodes.

For subsequent iterations $t > 0$, $A_L$ re-examines the literature to extract new snippets to update the hub and improve the current draft:
\begin{equation}
    \{F_1^{(t)}, F_2^{(t)}, \ldots, F_{N'}^{(t)}\} = A_L(\{\mathcal{L}_i\}_{i=1}^{I_k}, M^{(t-1)}_k)
\end{equation}
where $M^{(t-1)}_k$ denotes the draft content from the previous iteration, $\{F^{(t)}_n\}_{n=1}^{N'}$ is the set of newly extracted note snippets, and $N'$ is the number of snippets at iteration $t$. $A_D$ refines the section draft by incorporating the new evidence and the previous draft:
\begin{equation}
    M^{(t)}_k = A_D(\{F^{(t)}_n\}_{n=1}^{N'}, \mathcal{O}_k, M^{(t-1)}_k)
\end{equation}
where $M^{(t)}_k$ is the updated content, and $M^{(t-1)}_k$ serves as the previous context for refinement.

The iterative process for each outline node $\mathcal{O}_k$ terminates when the following condition is met:
\begin{equation}
    \psi(M^{(t)}_k, M^{(t-1)}_k) > \epsilon \quad \text{or} \quad t = t_{\max}
\end{equation}
where $\psi(\cdot, \cdot)$ denotes the similarity between consecutive iterations of the draft, measured by the ROUGE-1 score~\cite{lin2004rouge}, $\epsilon$ is a similarity threshold that indicates no further improvements can be made, and $t_{\max}$ is the maximum number of allowed iterations.

The final content $M^{(t^*)}_k$, where $t^*$ denotes the stopping iteration for outline node $\mathcal{O}_k$, serves as the final draft. Finally, $A_M$ aggregates the finalized drafts for all sections and delivers the completed literature review to the user.

\subsection{Self-Loop Memory}

The Self-Loop Memory is designed to execute tasks assigned by $A_M$, taking into account the risk of deviations at each step arising from unpredictable agent-environment interactions.

In detail, for each worker agent $A_W$ (which can be $A_S$, $A_O$, $A_L$, or $A_D$), $A_M$ initiates the Self-Loop Memory in the form of historical experience. $A_W$ first generates the primary results, then $A_M$ leverages historical experience to review these results, after which $A_W$ refines its outputs based on the feedback provided. At the initial step, when the historical experience is empty, $A_W$ produces the output solely based on the context:
\begin{equation}
    Y^{(0)} = A_W(C)
\end{equation}
where $Y^{(0)}$ is the initial output, and $C$ is the context fetched by $A_M$.

In the following steps, let $\mathcal{R}$ denote the set of all historical feedback records. For each feedback $r \in \mathcal{R}$, let $\text{revision}(r)$ denote the number of failures and further revisions required after applying feedback $r$, and let $\text{usage}(r)$ denote the number of times feedback $r$ has been selected for demonstration. Let $r_{\min} = \min_{r \in \mathcal{R}} \text{revision}(r)$ denote the minimum number of revisions across all feedback records. We select useful feedback and summarize it into a checklist to guide the worker's output at each iteration $x$:
\begin{equation}
    r^* = \arg\max_{r \in \mathcal{R}_{\min}} \text{usage}(r)
\end{equation}
where $\mathcal{R}_{\min} = \{ r \in \mathcal{R} : \text{revision}(r) = r_{\min} \}$ is the subset of feedback records that led to the fewest further revisions. Within $\mathcal{R}_{\min}$, we select the most frequently used one, i.e., the one with the highest $\text{usage}(r)$.

Referring to $r^*$, $A_M$ then provides feedback on the worker's output:
\begin{equation}
    R^{(x)} = A_M(Y^{(x)}, r^*)
\end{equation}
where $R^{(x)}$ denotes the feedback at iteration $x$.

For subsequent iterations ($x \geq 1$), $A_W$ refines the output with the input and the feedback:
\begin{equation}
    Y^{(x)} = A_W(C, R^{(x-1)})
\end{equation}

The Self-Loop Memory continues until the output meets the requirements of $A_M$ and no further feedback is provided, or until a maximum number of iterations $x_{\max}$ is reached.

In the following, we detail how worker agents cooperate with $A_M$. $A_S$ formulates search queries based on user instructions, submits queries to search engines, and collects relevant literature. It filters and ranks the retrieved references by citation count, removes incomplete entries, and selects the top $Q_k$ unique references for further use. $A_O$ utilizes the gathered literature and user instructions to prompt an LLM for generating sub-outlines. $A_L$ downloads and parses full-text papers, applies LLM-based previewing to select pertinent pages, and extracts concise text snippets for drafting. $A_D$ prompts the LLM to synthesize coherent text by incorporating selected references and prior drafts.

\begin{table*}[t]
    \centering
    \begin{adjustbox}{width=0.8\textwidth}
        \begin{tabular}{clllllll}
            \toprule
            \multirow{2.5}{*}{Survey Length} & \multirow{2.5}{*}{Methods} & \multicolumn{2}{c}{Citation Quality} & \multicolumn{4}{c}{Content Quality} \\
            \cmidrule(lr){3-4} \cmidrule(lr){5-8}
            & & Recall $\uparrow$ & Precision $\uparrow$ & Coverage $\uparrow$ & Structure $\uparrow$ & Relevance $\uparrow$ & Avg. $\uparrow$ \\
            \midrule
            \multirow{4}{*}{8k} 
            & DeepSeek-R1 (2025)   & 80.81 & 68.98 & 4.85 & 4.68 & 4.82 & 4.78 \\
            & Gemini-3 (2025)      & 85.93 & 86.35 & 4.93 & 4.88 & 4.94 & 4.92 \\
            & GPT-5.2 (2025)       & 87.12 & 76.67 & 4.95 & 4.82 & 4.92 & 4.90 \\
            & Naive RAG (2024)     & 78.14 & 71.92 & 4.40 & 3.86 & 4.86 & 4.37 \\
            & AutoSurvey (2024)    & 82.48 & 77.42 & 4.60 & 4.46 & 4.80 & 4.62 \\
            & SurveyX (2025)       & 85.72 & 78.35 & 4.93 & 4.92 & 4.96 & 4.94 \\
            & \cellcolor{gray!15}Proposed SGMC & \cellcolor{gray!15}\textbf{98.17} & \cellcolor{gray!15}\textbf{89.28} & \cellcolor{gray!15}\textbf{4.97} & \cellcolor{gray!15}\textbf{4.95} & \cellcolor{gray!15}\textbf{5.00} & \cellcolor{gray!15}\textbf{4.97} \\
            \midrule
            \multirow{4}{*}{16k} 
            & Naive RAG (2024)     & 71.48 & 65.31 & 4.46 & 3.66 & 4.73 & 4.28 \\
            & AutoSurvey (2024)    & 81.34 & 76.94 & 4.66 & 4.33 & 4.86 & 4.62 \\
            & SurveyX (2025)       & 85.28 & 78.14 & 4.91 & 4.88 & 4.94 & 4.91 \\
            & \cellcolor{gray!15}Proposed SGMC & \cellcolor{gray!15}\textbf{98.11} & \cellcolor{gray!15}\textbf{88.52} & \cellcolor{gray!15}\textbf{4.98} & \cellcolor{gray!15}\textbf{4.95} & \cellcolor{gray!15}\textbf{4.98} & \cellcolor{gray!15}\textbf{4.97} \\
            \midrule
            \multirow{4}{*}{32k} 
            & Naive RAG (2024)     & 79.88 & 65.03 & 4.41 & 3.75 & 4.66 & 4.27 \\
            & AutoSurvey (2024)    & 83.14 & 78.04 & 4.73 & 4.26 & 4.80 & 4.60 \\
            & SurveyX (2025)       & 85.43 & 78.21 & 4.94 & 4.89 & 4.95 & 4.93 \\
            & \cellcolor{gray!15}Proposed SGMC & \cellcolor{gray!15}\textbf{98.03} & \cellcolor{gray!15}\textbf{87.92} & \cellcolor{gray!15}\textbf{4.98} & \cellcolor{gray!15}\textbf{4.93} & \cellcolor{gray!15}\textbf{4.98} & \cellcolor{gray!15}\textbf{4.96} \\
            \midrule
            \multirow{4}{*}{64k} 
            & Naive RAG (2024)     & 68.79 & 61.97 & 4.40 & 3.66 & 4.66 & 4.24 \\
            & AutoSurvey (2024)    & 82.25 & 77.41 & 4.73 & 4.33 & 4.86 & 4.64 \\
            & SurveyX (2025)       & 85.14 & 78.02 & 4.95 & 4.90 & 4.95 & 4.93 \\
            & \cellcolor{gray!15}Proposed SGMC & \cellcolor{gray!15}\textbf{97.27} & \cellcolor{gray!15}\textbf{87.90} & \cellcolor{gray!15}\textbf{5.00} & \cellcolor{gray!15}\textbf{4.91} & \cellcolor{gray!15}\textbf{4.98} & \cellcolor{gray!15}\textbf{4.96} \\
            \bottomrule
        \end{tabular}
    \end{adjustbox}
    \caption{Comparison of automatic survey generation methods across different survey lengths (measured in tokens) on the benchmark dataset. Higher scores indicate better performance.}
    \label{tab:comparision_experiments}
\end{table*}

\section{Experiments}

\subsection{Experimental Settings}

We conduct experiments on two benchmark datasets, AutoSurvey \cite{wang2024autosurvey} and SurveyEval \cite{wang2025llm}, as well as a self-constructed benchmark dataset, TopSurvey. For hyperparameters, we set $d_{\max}$, $t_{\max}$, and $x_{\max}$ to 4. The hyperparameter $\theta$ is set to 500. We use GPT-4.1 \cite{achiam2023gpt} for all agents. The evaluation comprises two categories of metrics. For citation quality, we adopt citation recall and citation precision as proposed by \cite{wang2024autosurvey}. Recall measures whether cited passages fully support all statements. Precision measures the proportion of relevant citations that support their corresponding statements. For content quality, following \cite{wang2024autosurvey,liang2025surveyx}, we use coverage, structure, and relevance, each rated by LLMs on a 5-point scale. Coverage assesses the extent to which the survey encapsulates all relevant aspects of the topic, ensuring comprehensive discussion on both central and peripheral topics. Structure evaluates the logical organization and coherence of sections and subsections, ensuring that they are logically connected. Relevance measures how well the content of the survey aligns with the research topic and maintains a clear focus. We do not filter out fractional scores, such as 4.5. We use GPT-4o \cite{achiam2023gpt} as the evaluation model for fair comparison \cite{liang2025surveyx}.

\subsection{Comparison Experiments}

\begin{figure}[h]
    \begin{center}
        \includegraphics[width=0.8\linewidth]{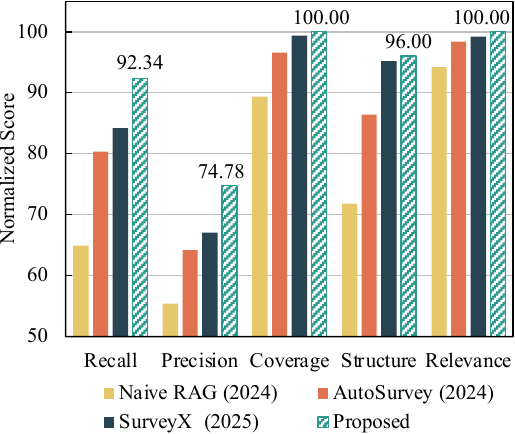}
    \end{center}
    \caption{Comparison of automatic survey generation methods on the SurveyEval benchmark. Higher scores indicate better performance.}
    \label{fig:surveyeval}
\end{figure}

\begin{table*}[h]
    \centering
    \begin{adjustbox}{width=0.6\linewidth}
        \begin{tabular}{lllllll}
            \toprule
            \multirow{2.5}{*}{Methods} & \multicolumn{2}{c}{Citation Quality} & \multicolumn{4}{c}{Content Quality} \\
            \cmidrule(lr){2-3} \cmidrule(lr){4-7}
            & Rec. $\uparrow$ & Pre. $\uparrow$ & Cov. $\uparrow$ & Str. $\uparrow$ & Rel. $\uparrow$ & Avg. $\uparrow$ \\
            \midrule
            Proposed SGMC w/o Tree-Guided  & 94.38 & 88.56 & 4.83 & 4.83 & 5.00 & 4.89 \\
            Proposed SGMC w/o Hub-Guided & 97.86 & 79.02 & 4.97 & 4.93 & 4.93 & 4.94 \\
            Proposed SGMC w/o Self-Loop & 97.78 & 80.82 & 4.79 & 4.89 & 4.97 & 4.88 \\
            \cellcolor{gray!15}Proposed SGMC & \cellcolor{gray!15}\textbf{98.17} & \cellcolor{gray!15}\textbf{89.28} & \cellcolor{gray!15}\textbf{4.97} & \cellcolor{gray!15}\textbf{4.95} & \cellcolor{gray!15}\textbf{5.00} & \cellcolor{gray!15}\textbf{4.97} \\
            \bottomrule
        \end{tabular}
    \end{adjustbox}
    \caption{Ablation study of the proposed modules on the benchmark dataset.}
    \label{tab:ablation_study}
\end{table*}

We first evaluate our framework on the AutoSurvey \cite{wang2024autosurvey} benchmark, following the protocols established by \cite{liang2025surveyx}. We use the same 20 topics from diverse subfields of LLM research to generate survey articles for comparison. We compare our approach with three baselines, including Naive RAG-based LLMs using retrieval-augmented generation, AutoSurvey \cite{wang2024autosurvey}, and SurveyX \cite{liang2025surveyx}. We follow the official report of AutoSurvey\footnote{https://github.com/AutoSurveys/AutoSurvey} and conduct experiments with the official implementation of the offline version of SurveyX\footnote{https://github.com/IAAR-Shanghai/SurveyX}.

As shown in Table \ref{tab:comparision_experiments}, most methods exhibit reduced performance as survey length increases, particularly in recall, precision, structure, and relevance. This decline is most pronounced in the Naive RAG baseline, indicating that longer workflows lead to greater error accumulation and propagation. In contrast, coverage remains stable or improves with longer surveys, likely due to more comprehensive topic inclusion. State-of-the-art methods such as AutoSurvey and SurveyX continue to face challenges with citation precision, frequently generating statements unsupported by references. For content quality, structure consistently receives the lowest scores, reflecting disorganized article organization. Our proposed method remains robust to these issues and consistently achieves superior results across all metrics.

We also compare with recent state-of-the-art LLMs, including DeepSeek-R1 \cite{guo2025deepseek}, Gemini-3 \cite{team2023gemini}, and GPT-5.2. These recent LLMs possess more up-to-date knowledge and enhanced reasoning capabilities. We equip them with an academic search tool, enabling them to retrieve relevant literature when necessary for generating the review. Even with paid subscriptions, state-of-the-art LLMs do not consistently generate long-form content exceeding 8k tokens as instructed, consistent with recent observations on the length control challenge \cite{akinfaderin2025plan}. Therefore, we report results only at the 8k-token length. The results show that Gemini-3 and GPT-5.2 demonstrate comparable performance to the agent system with GPT-4.1, highlighting the importance of the underlying foundation model.

We further evaluate our framework on the SurveyEval benchmark \cite{wang2025llm}, using the same protocols. SurveyEval is the first benchmark in computer science that pairs surveys with complete reference papers, comprising 384 arXiv cs.CL surveys citing over 26,000 references. Twenty topics are selected for testing based on reference completeness and reference list diversity. Experimental results are shown in Fig. \ref{fig:surveyeval}, with coverage, structure, and relevance scores normalized to a 100-point scale.

\begin{table*}[]
    \centering
    \begin{adjustbox}{width=0.75\linewidth}
        \begin{tabular}{lllllll}
            \toprule
            \multirow{2.5}{*}{Methods} & \multicolumn{2}{c}{Citation Quality} & \multicolumn{4}{c}{Content Quality} \\
            \cmidrule(lr){2-3} \cmidrule(lr){4-7}
            & Recall $\uparrow$ & Precision $\uparrow$ & Coverage $\uparrow$ & Structure $\uparrow$ & Relevance $\uparrow$ & Avg. $\uparrow$ \\
            \midrule
            Human                & 93.07 & 87.76 & 5.00 & 4.97 & 5.00 & 4.99 \\
            Naive RAG (2024)     & 64.57 & 61.89 & 4.29 & 3.58 & 4.67 & 4.18 \\
            AutoSurvey (2024)    & 70.03 & 71.66 & 4.68 & 4.67 & 4.87 & 4.74 \\
            SurveyX (2025)       & 75.51 & 77.90 & 4.71 & 4.84 & 4.93 & 4.83 \\
            \cellcolor{gray!15}Proposed SGMC & \cellcolor{gray!15}\textbf{86.63} & \cellcolor{gray!15}\textbf{81.98} & \cellcolor{gray!15}\textbf{4.85} & \cellcolor{gray!15}\textbf{4.90} & \cellcolor{gray!15}\textbf{5.00} & \cellcolor{gray!15}\textbf{4.92} \\
            \bottomrule
        \end{tabular}
    \end{adjustbox}
    \caption{Comparison of automatic survey generation methods at a survey length of 64k tokens on the new large-scale benchmark dataset. Higher scores indicate better performance.}
    \label{tab:comparison_64k}
\end{table*}

\begin{table*}[]
    \centering
    \begin{adjustbox}{width=0.85\linewidth}
    \begin{tabular}{cllllll}
    \toprule
    \multirow{2.5}{*}{Cluster} & \multicolumn{2}{c}{Citation Quality} & \multicolumn{4}{c}{Content Quality} \\
    \cmidrule(lr){2-3} \cmidrule(lr){4-7}
    & Recall $\uparrow$ & Precision $\uparrow$ & Coverage $\uparrow$ & Structure $\uparrow$ & Relevance $\uparrow$ & Avg. $\uparrow$ \\
    \midrule
    Physics & 82.94 & \textbf{81.43} & 4.69 & 4.76 & 4.91 & 4.79 \\
    Basic Medicine & 83.87 & 76.35 & 4.73 & 4.83 & 4.84 & 4.80 \\
    Education & 73.52 & 75.68 & 4.61 & 4.85 & 4.84 & 4.77 \\
    Computer Science \& Technology & \textbf{85.78} & 81.26 & \textbf{4.82} & \textbf{4.88} & \textbf{4.97} & \textbf{4.89} \\
    Economics & 80.91 & 71.24 & 4.68 & 4.75 & 4.76 & 4.73 \\
    \bottomrule
    \end{tabular}
    \end{adjustbox}
    \caption{Real-world case study across different topics on 10,000 generated reviews. Higher scores indicate better performance.}
    \label{tab:cluster_comparison}
\end{table*}

In Fig. \ref{fig:surveyeval}, all methods achieve high coverage and relevance, indicating that LLMs like GPT-4.1 can generate comprehensive and relevant content. However, recall and precision remain low, reflecting poor reference retrieval and insufficient support for generated statements. As a result, state-of-the-art methods struggle with logical organization, leading to lower structure scores. Our method overcomes these issues, achieving the best performance across all metrics.

\subsection{Ablation Study}
\label{sec:ablation_study}

We conduct an ablation study to evaluate the effectiveness of each component in our proposed framework. Experiments are performed on the same benchmark dataset as before, with all experimental settings unchanged and the survey length set to 8k tokens. We compare the full model with three variants. For the variant without Tree-Guided Memory, we use a single round of retrieval and organization to generate the overall outline. For the variant without Hub-Guided Memory, we draft the manuscript using only one round of extraction and writing. For the variant without Self-Loop Memory, each agent completes its assigned task without revision.

Table \ref{tab:ablation_study} presents the results. Removing Tree-Guided Memory causes the largest drops in recall and structure, demonstrating its importance for citation recall and logical organization. Excluding Hub-Guided Memory sharply reduces precision and relevance, confirming its role in improving citation precision and content relevance. Without Self-Loop Memory, all metrics decrease, especially coverage, highlighting its key role in ensuring comprehensive coverage through collaborative revision.

\subsection{A New Large-Scale Benchmark}

To further validate the effectiveness of our proposed framework, we construct a new large-scale benchmark dataset. This dataset consists of 195 topics from various subfields of computer science, nearly 10 times larger than previous benchmarks \cite{wang2024autosurvey,liang2025surveyx}. To ensure topic quality, we collect peer-reviewed survey topics from top computer science conferences, rather than from preprint sources such as arXiv. Survey papers accepted only as abstracts are excluded. To prevent data leakage from LLM pretraining data, we include only survey papers published in 2023, 2024, and 2025. We employ PhD students in computer science to verify whether a paper qualifies as a survey and to collect the final set of 195 survey papers: 9 from AAAI, 34 from ACL, 46 from EMNLP, 3 from ICLR, 2 from ICML, 77 from IJCAI, and 24 from NAACL. Of these, 68 were published in 2023, 105 in 2024, and 22 in 2025.

We conduct experiments using the same settings as above, generating 64k-token literature reviews for evaluation. As shown in Table \ref{tab:comparison_64k}, compared with the results on the existing benchmark in Table \ref{tab:comparision_experiments}, performance on the new large-scale benchmark drops significantly, particularly in recall and coverage. This may be due to the greater number and broader range of topics, which leads to some relevant literature not being retrieved, resulting in lower citation recall and coverage. Despite this, our proposed method achieves over 80\% in citation scores and an average content quality score of 4.92.

Finally, we evaluate our framework in a real-world setting by deploying an online automatic literature review generation system\footnote{Anonymous for review; to be revealed upon acceptance.}, which has produced approximately 40,000 reviews to date. We randomly sampled 10,000 reviews, embedded them using all-MiniLM-L6-v2, and clustered them into five groups corresponding to the five most frequently requested disciplines. As shown in Table \ref{tab:cluster_comparison}, results reveal notable differences in citation and content quality across clusters. The best performance is observed in Computer Science \& Technology, while Education performs comparatively worse, particularly in citation quality. This discrepancy may be attributed to the search engine's limited coverage of relevant literature in the social sciences. We also measured system efficiency: generating an 8,000-token review takes an average of 8.45 minutes. To facilitate future research, we release the titles and instructions of the 10,000 sampled reviews in English for comparative studies.

\section{Discussion}

Our experimental results demonstrate that SGMC outperforms not only pipeline-based agents but also the latest standalone LLMs equipped with retrieval tools. This raises a crucial question: in the era of increasingly powerful foundation models, what is the unique value of a structured framework like SGMC? Our structured memory mechanism offers two key advantages over pure LLM-based approaches. First, the tree-guided outline and hub-guided memory nodes create an explicit, traceable chain from any final claim back to its source literature, enabling errors to be pinpointed to specific retrieval, outlining, or drafting steps. Standalone LLMs lack this transparent provenance, making error diagnosis and systematic correction unreliable. Second, SGMC embeds proactive verification within the workflow through Self-Loop Memory, preventing errors from propagating to subsequent stages. In contrast, methods applicable to standalone LLMs, such as self-refinement or post-editing, are inherently reactive and often insufficient for correcting deeply embedded, compounded errors. The ablation study further substantiates that each component of our structured memory is vital for holistic quality. We acknowledge the remarkable capabilities of state-of-the-art LLMs, and the agents within SGMC can be instantiated by these powerful models. The contribution of SGMC is to provide a reliable, controllable, and error-aware process framework that channels the generative power of LLMs toward producing high-fidelity, verifiable scholarly content. For automated literature review generation, this structural guidance is essential.

\section{Conclusion}

In this paper, we tackle the challenge of compounding errors in automated long-form literature review generation. We propose Structure-Guided Memory Consolidation (SGMC), a framework that systematically consolidates and verifies information throughout the multi-stage generation process using tree-guided, hub-guided, and self-loop memory modules. Extensive experiments demonstrate that SGMC achieves state-of-the-art performance in citation accuracy and content quality. Moreover, the successful deployment of our online system, which has generated over 40,000 reviews, provides strong evidence of SGMC's practical effectiveness.

\bibliographystyle{named}
\bibliography{ijcai26}

\end{document}